\documentclass[epjST]{svjour}
\usepackage{url,color,colordvi,epsfig,pstricks}
\usepackage[colorlinks=true, citecolor=blue, filecolor=blue, linkcolor=blue, urlcolor=blue, pdftex]{hyperref}
\usepackage{graphicx,bm,dcolumn}
\usepackage{hyperref,slashed,lineno}
\usepackage{physics}
\usepackage{inputenc}
\usepackage{ulem}
\usepackage[square,numbers]{natbib}

\hypersetup{
    colorlinks=true, 
    linktoc=all, 
    linkcolor=blue, 
    citecolor=blue}

\begin{document}

\title{Nuclear medium effects in neutrino- and antineutrino-nucleus scattering}
\author{N. Jachowicz\inst{1}\thanks{natalie.jachowicz@ugent.be} \and A.~Nikolakopoulos\inst{1}\inst{2}\thanks{alexis.nikolakopoulos@ugent.be}}
\institute{Ghent University, Department of Physics and Astronomy, Proeftuinstraat 86, B-9000 Gent, Belgium \and University of Geneva, Section de Physique, DPNC, Geneva, Switzerland}

\abstract{
In this paper we study the influence of nuclear medium effects on quasi-elastic neutrino-nucleus scattering processes. We focus on effects provided by the nuclear mean field and random phase correlations and pay special attention to differences between neutrino- and antineutrino-induced reactions. We confront our results with the T2K and MiniBooNE data for both neutrinos and antineutrinos and the neutrino-antineutrino asymmetry.  In view of the recently published ab initio results we provide a careful comparison between our cross section predictions and the ab-initio calculations.
}

\maketitle

\section{Introduction}
One of the most exciting physics searches of this decade is the one seeking to unravel  the role of neutrinos for CP violation in weak interactions.  Current neutrino oscillation experiments are preparing for a final quest  in determining the CP violating phase delta, controlling the differences in behavior between particles and antiparticles in the weak sector. In oscillation experiments, tiny differences in oscillation characteristics for neutrinos and antineutrinos will be the signalers of this CP violation and allow to extract a value for $\delta$ from the experiment.  That this is a challenging task is made evident by the current status of the field : the T2K and NOvA collaborations recently reported strongly differing results with regards to the value of this parity violating delta phase \cite{T2K:CP2020, NOVAPRL2019}.

One of the major uncertainties in these measurements and in neutrino-oscillation experiments in general, is related to the role of nuclear effects on the observed cross sections~\cite{NUSTECWP,KatoriMartinireview}.  In accelerator-based oscillation experiments, nuclei are the most important detection material.  Whereas this results in a welcome enhancement of the detection signal, this complicates its interpretation as effects induced by the nuclear medium strongly influence the signature the weak interaction leaves in the detector.

As the first generation of accelerator-based oscillation experiments provided only fully inclusive data, several models were developed in an inclusive fashion, to deal with cross section calculations in an numerically efficient way. Superscaling methods \cite{PauliBlocking,Amaro:2018xdi,SuSAMstarneutrino} e.g.~are based on the scaling behaviour exhibited by electron scattering data, ported to  the weak sector by making use of relativistic mean-field (RMF) modeling. SuSAv2 in particular is further extended to explicitly include meson-exchange contributions to the cross-sections \cite{SuSAMEC,Megias:2017cuh,DePace:2003spn} and the inelastic response of the nucleus~\cite{BostedChristy,SuSAJlabAr}. Several models \cite{Nieves:2017lij, MartiniModel:2009, MartiniModel:2010} are based on Fermi-gas approaches made more realistic by including random-phase-approximation (RPA)-like correlations and spectral-function corrections.
Spectral function approaches~\cite{BENHAR2005,BENHAR2010,BENHAR2015,Rocco16,Ankowski15a,Rocco:2019gfb} exploit the (plane-wave) impulse approximation to factorize cross sections and include the influence  of a broad range of medium effects that can be encoded in the (effective) spectral function.
Other models are based on the relativistic distorted-wave impulse approximation (RDWIA). These generally use a relativistic mean field (RMF) shell model to describe the initial state~\cite{Horowitz81}, possibly modified by a spectral function~\cite{Gonzalez-Jimenez:2021ohu}, and a suitable potential to describe the scattered nucleon wave function ~\cite{Giusti2012MB,Gonzalez-Jimenez13c,Gonzalez-Jimenez19}.
The relativistic Green's function technique, closely related to the RDWIA, in particular aims to consistently treat the inclusive scattering cross section with an optical potential suitable to describe the semi-inclusive final-state~\cite{PhysRevC.84.015501,Ivanov16b}.
Recently the Green's function Monte Carlo method~\cite{Lovato:PRC91} was used to calculate the inclusive electroweak response functions of carbon over a kinematic range suitable to describe flux-averaged neutrino-nucleus cross sections~\cite{Lovato:PRX}.
This approach uses what is essentially a closure relation to integrate over all final-state configurations, such that only inclusive responses are attainable.
The short-time approximation method as presented in Ref.~\cite{Pastore:2019urn} aims to retain the main appeal of ab-initio methods, namely the consistent treatment of one- and two-body contributions, while providing information on the hadronic final state. 

In the following paragraphs, we will present our cross section results obtained in a Hartree-Fock continuum random phase approximation (HF-CRPA) approach for inclusive neutrino and antineutrino-induced quasi-elastic processes and confront them with data and with the results of ab-initio calculations.

\section{Quasi-elastic (like) cross sections}
The cross sections we present in this work are based on a  mean-field framework, supplemented with long-range correlations obtained in a continuum random phase approximation (CRPA) approach~\cite{RYCKEBUSCH1988,CRPAmod,JachoNC}.  In a mean field description, the nucleon is subject to nuclear correlations, summarized in the averaged field provided by the interactions with all the other nucleons in the medium and limited to correlations included in the Hartree-Fock propagator.   The initial  and final-state Hartree-Fock nucleon wave functions are  obtained as solutions of the Schrodinger equation with a Hartree-Fock (HF) potential generated from the SkE2 Skyrme effective parameterization of the nucleon-nucleon interaction \cite{WAROQUIER1983}. This description results in a Slater determinant description of the nucleus, with a shell structure in which nucleons populate orbitals with well determined angular momentum quantum numbers and energy.  Pauli blocking, binding for the nucleons in different shells and elastic distortion of the outgoing nucleon's wave function are taken into account in a straightforward manner.
In the CRPA approach, long-range correlations are  included in a consistent way using the same SkE2 parameterization used for the mean-field potential as residual interaction.   CRPA transition densities are obtained as superposition of contributions of  particle-hole (ph) and hole-particle excitations out of a correlated ground state. In this way, the description goes beyond a spectator approach i.e. the nucleon interacting with the neutrino at the weak vertex will still exchange energy with the other nucleons before it eventually leaves the nucleus. In our description, the RPA formalism is formulated in a propagator description, and the CRPA equations are solved in coordinate space, allowing to take the continuum into account in a full way without need to introduce a discretization scheme or upper bounds. For low excitation energies in particular, the RPA allows to describe the strength provided by collective excitations in the giant resonance region. While the separation between the mean-field, e.g. the Hartree-Fock nucleus in this case, and 'correlations' is in principle formally clear, the exact physics content of the mean field and various types of correlations is model-dependent, and determined by the content of the nucleon-nucleon interaction.

\begin{figure}
    \includegraphics[width=\textwidth]{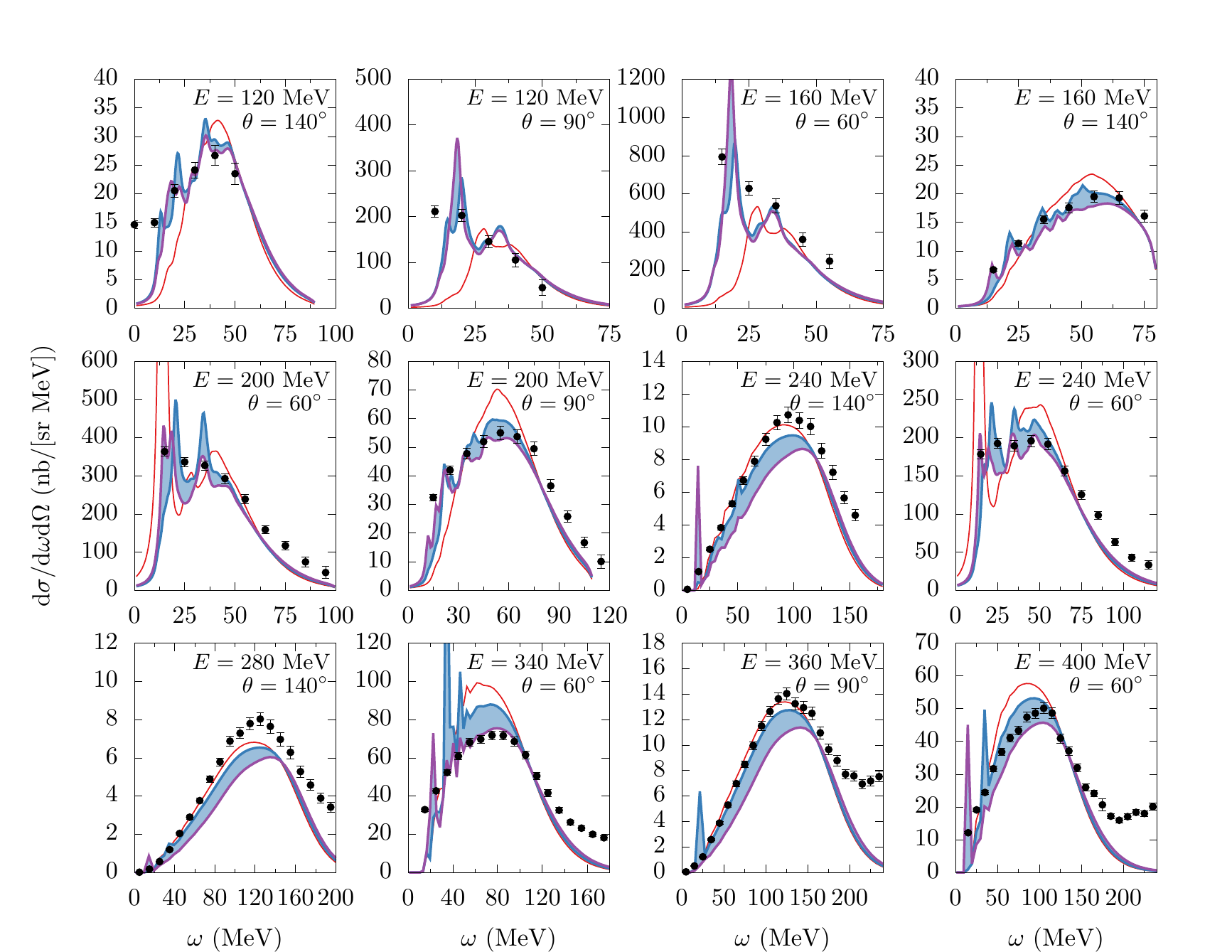}
    \caption{
    Comparison of Hartree-Fock (red lines) and CRPA (blue band) results for inclusive electron scattering data off iron. The blue line corresponds to the CRPA model with a dipole cut-off in the residual interaction (see text), while in the purple line no cut-off is included. The data is taken from Refs.~\cite{Saclay:CA,ee-data}}
    \label{fig:eeprime_iron}
\end{figure}

Relativistic effects are taken into account along the lines provided by the prescription outlined in Ref.~\cite{Jeschonnek}. 
Coulomb distortion of the outgoing lepton in the potential of the residual nucleus is implemented with an interpolation scheme combining a low and a higher energy approach. For small lepton energies the correction is done by multiplying the cross section with the appropriate form of the Fermi function. As this correction factor equals the ratio of the s-wave  solution for the lepton wave-function evaluated in the Coulomb potential of the nucleus to the plane wave solution, it is not valid for larger lepton energies.  For higher energies, the correction is provided by the modified effective momentum approximation (MEMA) \cite{MEMA}, where the momentum of the final lepton is shifted to an effective value by correcting its value with the Coulomb energy evaluated at the center of the nucleus, and modifying the phase space representing the density of final states accordingly.  At each kinematics, the scheme providing the smallest correction is selected as explained in Refs.~\cite{VanDessel:2019atx,VanDessel:2019obk}.

As the CRPA calculation is based on an effective interaction fitted to ground-state and low-lying excited state properties of a range of nuclei, it lacks the right $Q^2$ behavior to describe long-range correlations  in an accurate way for reactions at higher energy and momentum transfer and provides far too much strength at these higher kinematics.  Therefore in our calculations of neutrino-nucleus cross sections the high $Q^2$ behavior of the interaction can be constrained by a dipole behavior, with a cut-off mass fitted to a broad range of electron-scattering data \cite{PRC92}.  

Short-range correlations (SRC) and meson-exchange currents (MEC) give rise to corrections to 1-nucleon knockout cross sections and provide supplementary 2-nucleon (2p2h) knockout strength.  We studied the influence of SRCs in neutrino scattering within the Hartree-Fock framework in \cite{VanCuyck:2016} and of seagull currents in Ref.~\cite{VanCuyck:2017wfn}, work on the further extension of MECs is in progress. In this work we will use the SuSAv2 2p2h predictions \cite{SuSAMEC} when relevant for the comparison with data.

\section{Inclusive electron scattering}
While the dataset for neutrino-nucleus interactions in the several-hundred MeV and few-GeV regions has seen significant expansion in recent years, it is currently not yet possible to clearly constrain models for a specific interaction mechanism. This is both due to limited statistics, but more-so due to the fact that neutrino beams span a wide energy range such that the experimental data is comprised out of several different interaction mechanisms which are not clearly separable, and which are most often not described within a consistent model. 
 It is hence a prerequisite of any nuclear model used in the calculation of neutrino-nucleus scattering, to benchmark the vector part of the response against electron scattering data~\cite{Amaro:2020enu}.   The results of the HF-CRPA model have been compared with electron scattering data for several target nuclei in Refs.~\cite{Pandey:2016, Nikolakopoulos:KDAR}.
With a growing number of neutrino experiments aiming to use heavier, and often isospin asymmetric nuclei as target material, here we make a comparison of our HF-CRPA cross section predictions with experimental electron scattering data off ${}^{56}$Fe  in Fig.~\ref{fig:eeprime_iron}.
We focus on the region of relatively low energy and momentum transfer, where nuclear structure effects can be expected to become increasingly important.

Once the nucleon-nucleon interaction is fixed, the scattering cross section as predicted in the HF model is essentially parameter-free, relying only on the input of the free nucleon form factors.
As mentioned before, in the CRPA, only one single free parameter is introduced, namely the cut-off mass in the residual interaction.
This cut-off mass used in the dipole fit for the residual CRPA interaction was obtained in a global fit to $(e,e^\prime)$ cross section data spanning a large kinematic range.
From the comparison in Fig~\ref{fig:eeprime_iron}, and as pointed out in Ref.~\cite{Nikolakopoulos:KDAR}, it is clear that whereas this single dipole cut-off may provide the best fit to a global set of electron-scattering data, in certain kinematic regimes, namely the region of small to intermediate energy and momentum transfer which is of importance for the T2K and MiniBooNE data considered in this work, the full CRPA description often gives a better description of the data.
The CRPA results in Fig.~\ref{fig:eeprime_iron} are therefore shown with and without the cut-off as the upper and lower bounds of an uncertainty band respectively. Overall the agreement between model and data is excellent. 
 At intermediate momentum transfers the RPA provides the required suppression with respect to the HF results.
  In the giant resonance region however, collective RPA excitations provide extra strength on top of the mean field contribution.  

\begin{figure}
    \includegraphics[width=\textwidth]{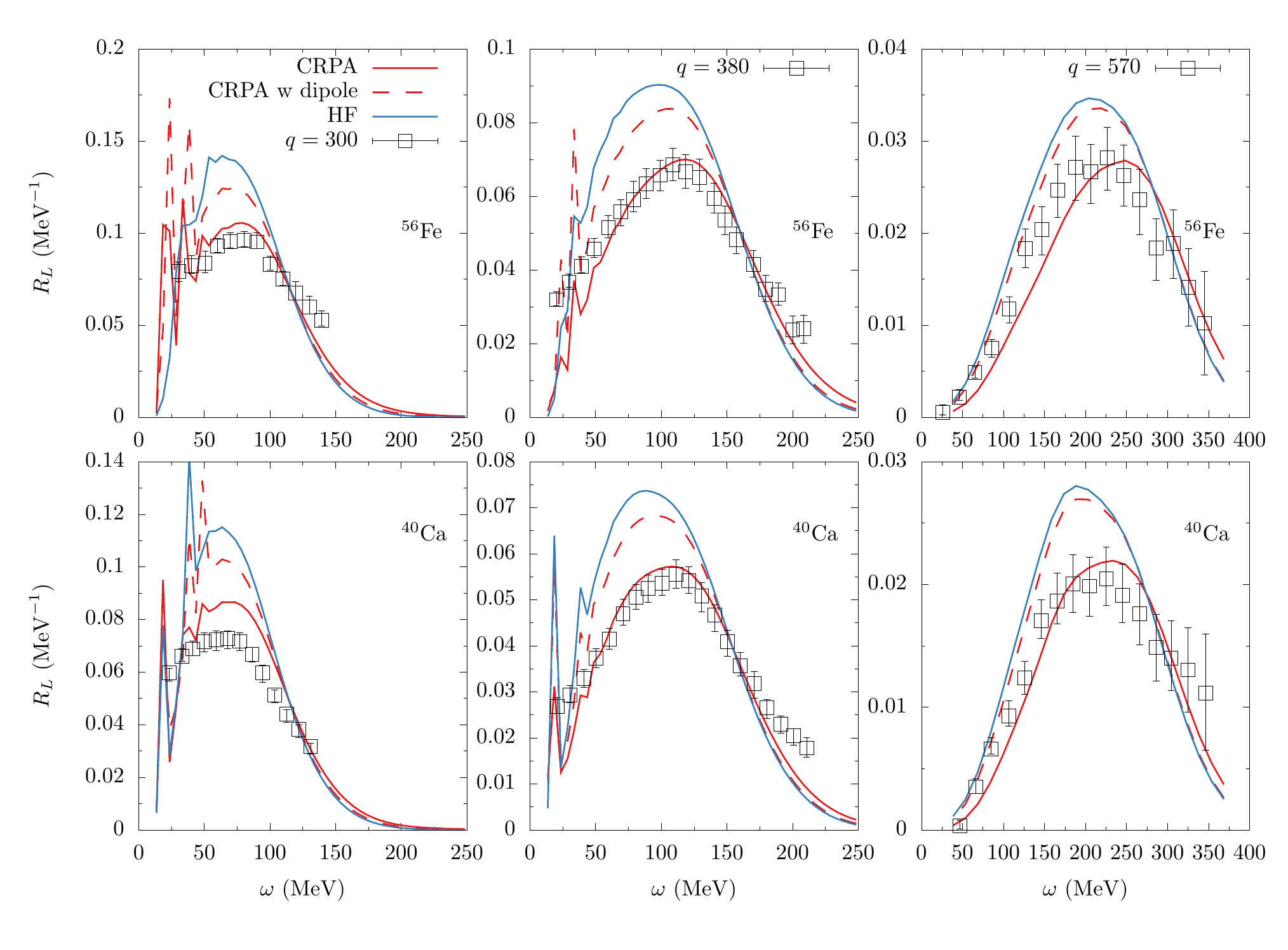}
    \caption{
        Comparison between HF and CRPA predictions and data for the  electromagnetic longitudinal response for calcium and iron. The data corresponds to the world-data analysis of Ref.~\cite{JOURDAN1996117}.}
    \label{fig:RL}
\end{figure}

\begin{figure}
    \includegraphics[width=\textwidth]{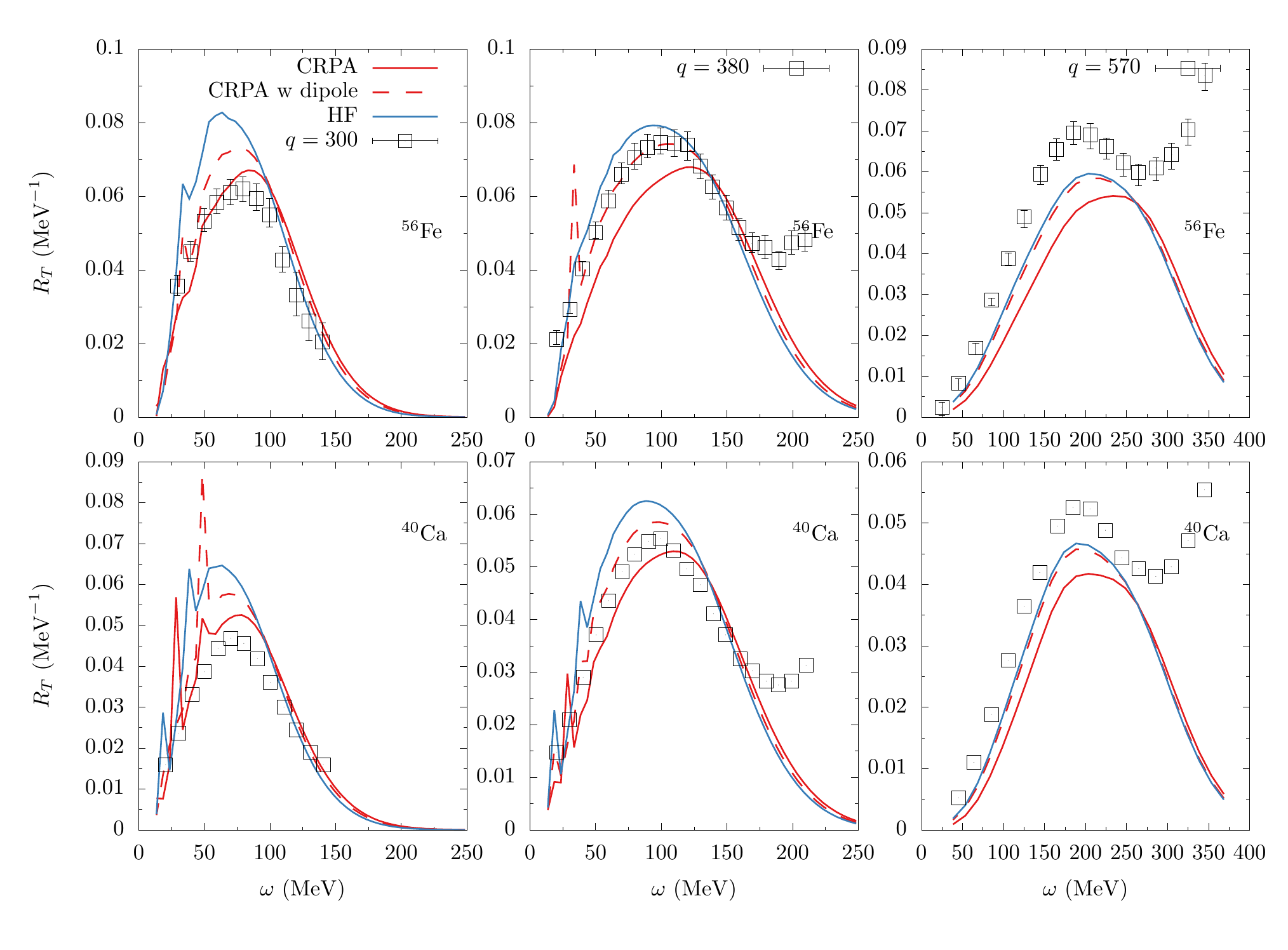}
    \caption{
    Same as in Fig.~\ref{fig:RL} but for the  transverse electromagnetic response.}
    \label{fig:RT}
\end{figure}

In Figs. \ref{fig:RL} and \ref{fig:RT} we present the comparison between Rosenbluth separated data and calculated longitudinal and transverse responses for calcium and iron at different values of the momentum transfer, a comparison to the carbon data was shown in Ref.~\cite{PRC92}.
The responses are obtained from the analysis of the world data of Ref.~\cite{JOURDAN1996117}.
We find an enhanced improvement in the description of data for the heavier nuclei considered here, in particular for the longitudinal response.  When the residual interaction is included with its full strength the CRPA provides a reduction of the longitudinal response, which is relatively stronger than in the transverse case, that brings the model closer to the data.
For the transverse case the overall magnitude of the CRPA reduction is smaller and mostly in line with the data, but it tends to introduce a shift towards higher values of $\omega$ when $q$ becomes large.
This correlates with the fact that the transverse response is more strongly affected by 2-body currents, of which the exact physics content depends on the nucleon-nucleon interaction, 
and where in particular e.g. delta-currents are  not included in the HF-CRPA results presented here.

While the separated responses should in principle provide a clear insight in the agreement with electron scattering data, free from the kinematic dependence introduced by the lepton vertex, we are hesitant to draw any conclusions from this comparison.
First because the effect of meson-exchange, and in particular delta currents is not taken into account in this comparison.
On top of this, there is some significant ambiguity in extracting the responses from electron scattering data.
This is partly due to the uncertainty from Coulomb effects, which are estimated usually with effective schemes~\cite{JOURDAN1996117} and are subsequently divided out of the experimental cross section in order to obtain the nuclear responses.  Moreover, in particular for the longitudinal response which provides a relatively small contribution to the cross section for more backward kinematics, the systematic error could be expected to be rather large~\cite{PhysRevC.45.1333}. This is clearly illustrated by the important differences that exist between two of the experimental datasets for calcium that are included in this data~\cite{JOURDAN1996117}; the Bates data of Ref.~\cite{Bates:CA40} and the Saclay dataset~\cite{Saclay:CA}. As a consequence the reported experimental uncertainty indicated by the error-bars in Figs.~\ref{fig:RL}~and~\ref{fig:RT} is likely to be far too optimistic, and the actual systematic error budget (especially for the longitudinal response) may well be far larger.

\section{Neutrino versus antineutrino induced interactions}
For  charged current interactions, the differential quasi-elastic cross section can be expressed as 
\begin{eqnarray}
    \frac{d^3 \sigma}{dE_f d^2\Omega_f}&=&
    \frac{G_{F}^2\cos^2\theta_c}{\pi}E_f k_f \zeta^2\\
  &&\times [v_{CC} W_{CC}+v_{CL}W_{CL}+v_{LL}W_{LL}+v_T W_T+hv_{T^\prime}W_{T^\prime}] \nonumber\\
\end{eqnarray}
with $E_f$, $k_f$, $\Omega_f$ denoting energy, momentum, and direction of the outgoing lepton, and $h$ the neutrino helicity. $G_F$ is the Fermi coupling constant, $\theta_c$ the Cabibbo angle. The factor $\zeta^2$ denotes the multiplicative Coulomb correction for the lepton as described in~\cite{VanDessel:2019atx}.  The $v$ and $W$ factors are representing the lepton kinematic factors and the nuclear responses, the latter encoding the HF-CRPA effects as described in the previous paragraph, expressions for both can be found in Ref.~\cite{Jachowicz:JPG2019}. 

\begin{figure}
    \centering
    \includegraphics[width=\textwidth]{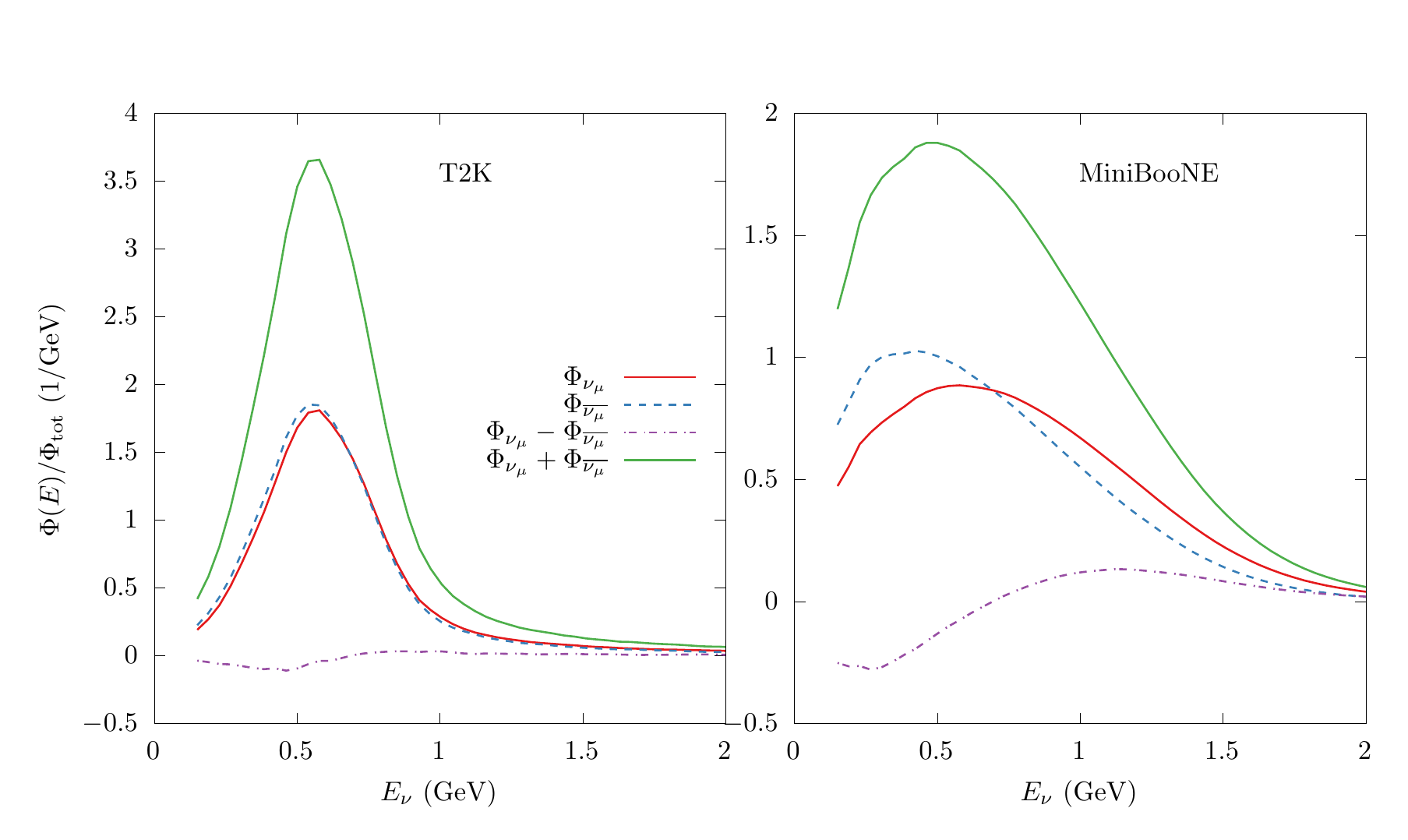}
    \caption{The MiniBooNE and T2K fluxes for neutrinos and antineutrinos normalized to one, along with their difference and sum. }
    \label{fig:fluxes}
\end{figure}

The combination of the vector-axialvector ($V-A$) structure of the weak interaction, strong isospin symmetry, and   (anti) neutrinos in practice  being (positive) negative helicity eigenstates, gives rise to the fact that for reactions off a nucleon, the difference between neutrino and antineutrino induced interactions is simply accounted for by a change of sign of the axial current.

In nuclei however there is no perfect symmetry between protons and neutrons.
This is clearly the case when $N > Z$, with the neutrons occupying additional single-particle orbitals.
But even if N=Z, as is the case for e.g.~carbon and oxygen.  Small differences in the proton and neutron cross sections arise from various effects induced by the nuclear charge~:  the Coulomb potential is responsible for the distortion of outgoing proton's wave function, bound protons and neutrons have different binding energies even for states with the same quantum numbers, and the bound proton wave functions are influenced by the Coulomb potential.
These effects are included in our mean-field calculations, even for symmetric nuclei.
Moreover, in charged current interactions, the Coulomb potential of the final outgoing lepton, treated effectively as outlined above, is attractive for negatively charged leptons and repulsive for positive leptons.

\begin{figure}
    \includegraphics[width=\textwidth]{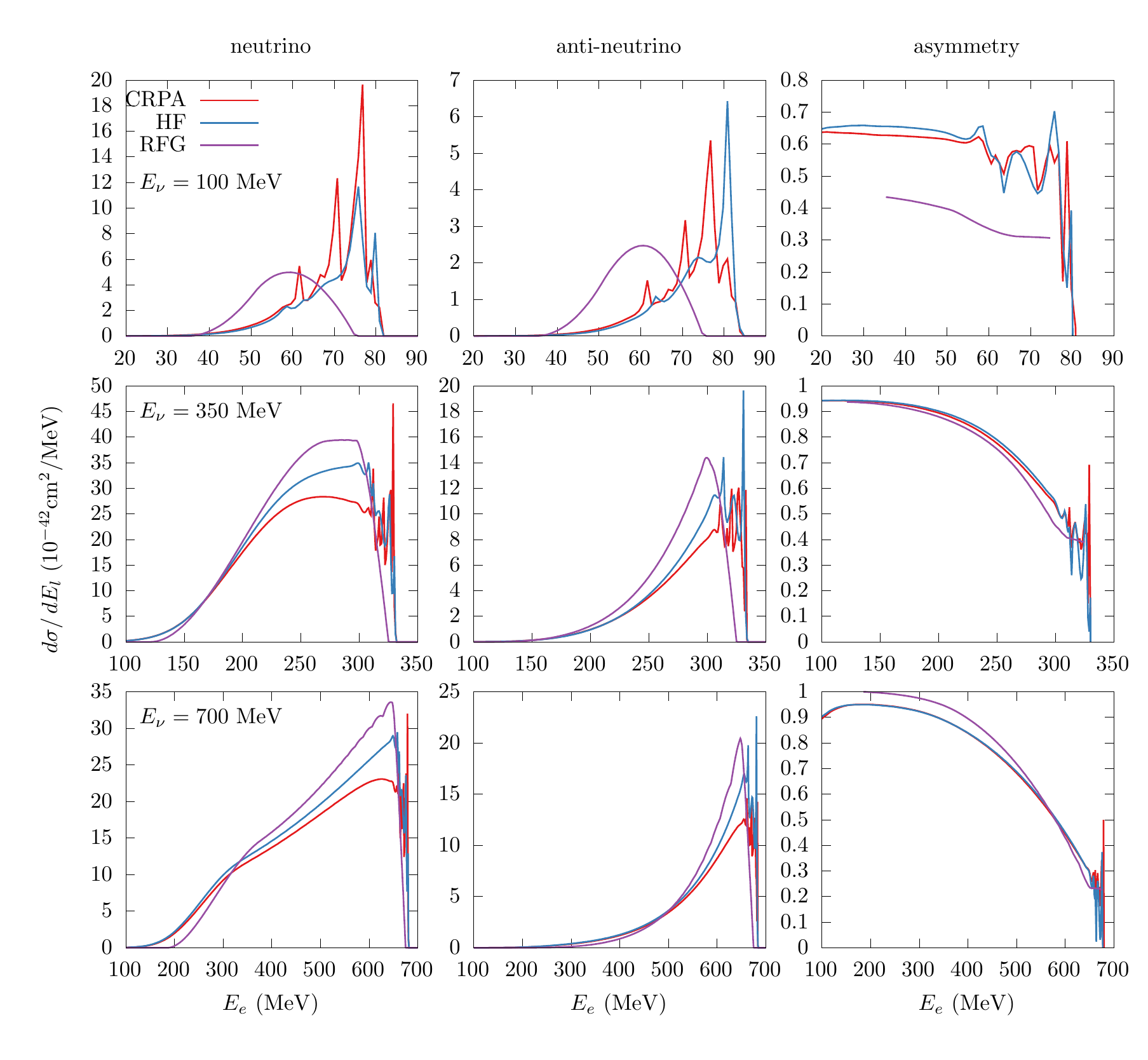}
    \caption{Angle-integrated differential cross sections for charged-current electron neutrino (left) antineutrino (middle) scattering and asymmetry (right) off $^{12}$C as a function of outgoing lepton energy, for incoming energies of 100 (top), 350 (middle) and 700 MeV (bottom). Different curves show the HF, CRPA and RFG results. The RFG results are obtained using the expressions in Ref.~\cite{Amaro:2005} and include a binding energy of $25~\mathrm{MeV}$.}
    \label{fig:simpel}
\end{figure}

\begin{figure}
    \includegraphics[width=\textwidth]{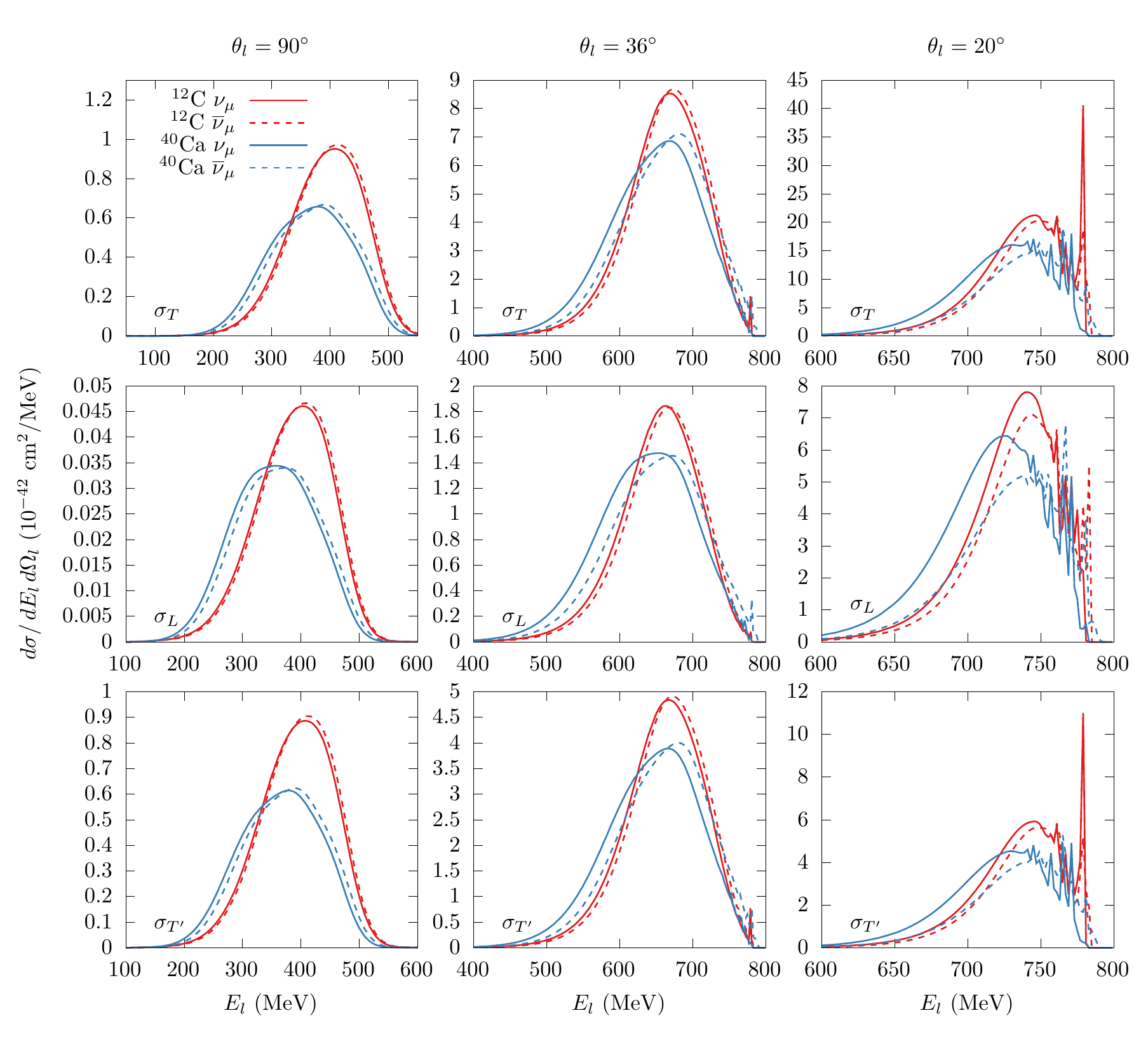}
    \caption{Cross sections per nucleon for muon neutrino (solid) and anti-neutrino (dashed) scattering off carbon (red) and calcium 40 (blue) targets at an incoming energy of 800 MeV.
    Top row shows transverse (T), middle  longitudinal (CC+CL+LL) and bottom is $T^\prime$ (VA interference) contributions to the cross section.
    The columns correspond to different fixed muon scattering angles.
    }
    \label{fig:DDIFfixEA}
\end{figure}

In the comparison to experimental data, variations between neutrino and antineutrino energy distributions are a further cause of differences between neutrino and antineutrino induced processes. Fig. \ref{fig:fluxes} compares normalized neutrino and antineutrino fluxes for MiniBooNE and T2K~\cite{MBflux:2009,T2K:flux2016}.  For T2K, shape differences between both fluxes are extremely small, for MiniBooNE, the energy distribution for neutrinos peaks at slightly higher energy than the antineutrino one and has a clearly more pronounced high-energy tail.

The neutrino-antineutrino cross section asymmetry can be defined as 
\begin{equation}
    A=\frac{\sigma_\nu - \sigma_{\overline{\nu}}}{\sigma_\nu + \sigma_{\overline{\nu}}},
\end{equation}
where $\sigma$ is shorthand for the relevant (flux-averaged) differential cross section.
From a purely theoretical perspective, and when small isospin breaking effects can be neglected, as e.g. for carbon or oxygen nuclei, one may write
$\sigma_\nu = \sigma_{VV,AA} + \sigma_{VA}$ and $\sigma_{\overline{\nu}} = \sigma_{VV,AA} - \sigma_{VA}$, such that for the asymmetry in flux-folded experimental data
\begin{eqnarray}
\label{eq:asymMB}
A &=& \frac{\int \Phi_\nu(E_\nu)\sigma_\nu (E_\nu)dE_{\nu}- \int \Phi_{\overline{\nu}}(E_{\overline{\nu}})\sigma_{\overline{\nu}}(E_{\overline{\nu}})dE_{\overline{\nu}}}{\int \Phi_\nu(E_\nu)\sigma_\nu (E_{\nu})dE_{\nu}+ \int \Phi_{\overline{\nu}}(E_{\overline{\nu}})\sigma_{\overline{\nu}}(E_{\overline{\nu}})dE_{\overline{\nu}}} \\
&=& \frac{\int\,dE \left( \Phi_\nu - \Phi_{\overline{\nu}} \right)\sigma_{VV,AA} + \left(\Phi_\nu + \Phi_{\overline{\nu}} \right)\sigma_{VA}}{\int\,dE \left( \Phi_\nu + \Phi_{\overline{\nu}} \right)\sigma_{VV,AA} + \left(\Phi_\nu - \Phi_{\overline{\nu}} \right)\sigma_{VA}}. \label{eq:iso}
\end{eqnarray}
This means that only when the shape of the neutrino and antineutrino fluxes is identical, the asymmetry selects the  vector-axial interference term in the numerator and the vector-vector plus axial-vector terms in the denominator as is the case for a fixed incoming energy under perfect isospin symmetry.
As Fig.~\ref{fig:fluxes} shows that in the T2K experiment the shape of the neutrino and antineutrino fluxes is very similar, hence their sum will be far larger than their difference in Eq. \ref{eq:asymMB}. If the very small difference is neglected, one obtains
\begin{equation}
\label{eq:fluxsm}
A = \frac{\int\,dE \left(\Phi_\nu + \Phi_{\overline{\nu}} \right)\sigma_{VA}}{\int\,dE \left( \Phi_\nu + \Phi_{\overline{\nu}} \right)\sigma_{VV,AA}}.
\end{equation}
We point out that in the following we make no such assumptions on the shape of the flux, nor on the isospin symmetry of the target, i.e. we use Eq.~\ref{eq:asymMB} for all calculations rather than Eqs.~\ref{eq:iso} or~\ref{eq:fluxsm}.

\begin{figure}
    \includegraphics[width=\textwidth]{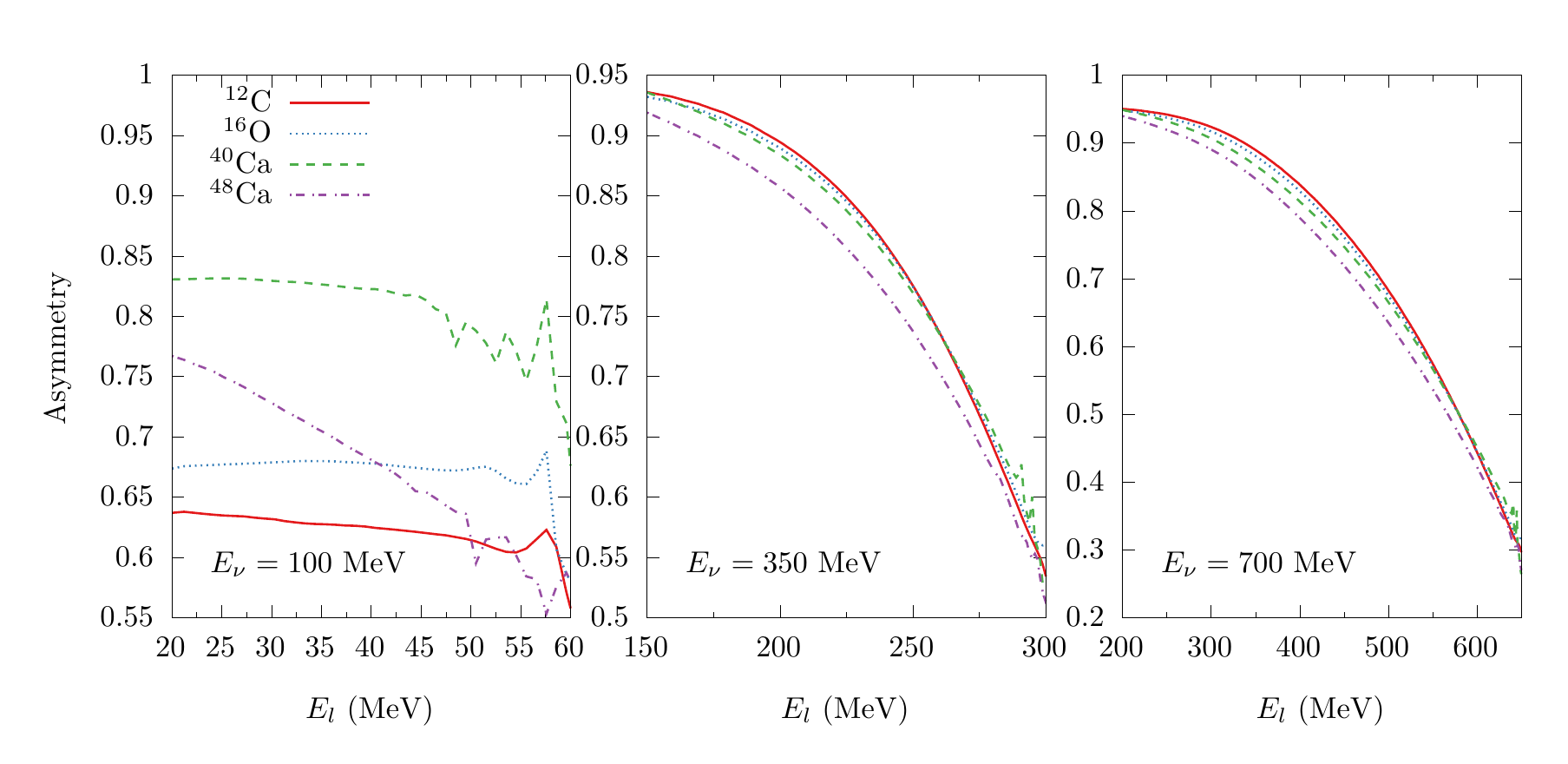}
    \caption{CRPA asymmetry for  symmetric even-even nuclei, and for $^{48}$Ca, corresponding to incoming energies of 100, 350, and 700 MeV. }
    \label{fig:simpelAdep}
\end{figure}

\begin{figure}
    \includegraphics[width=\textwidth]{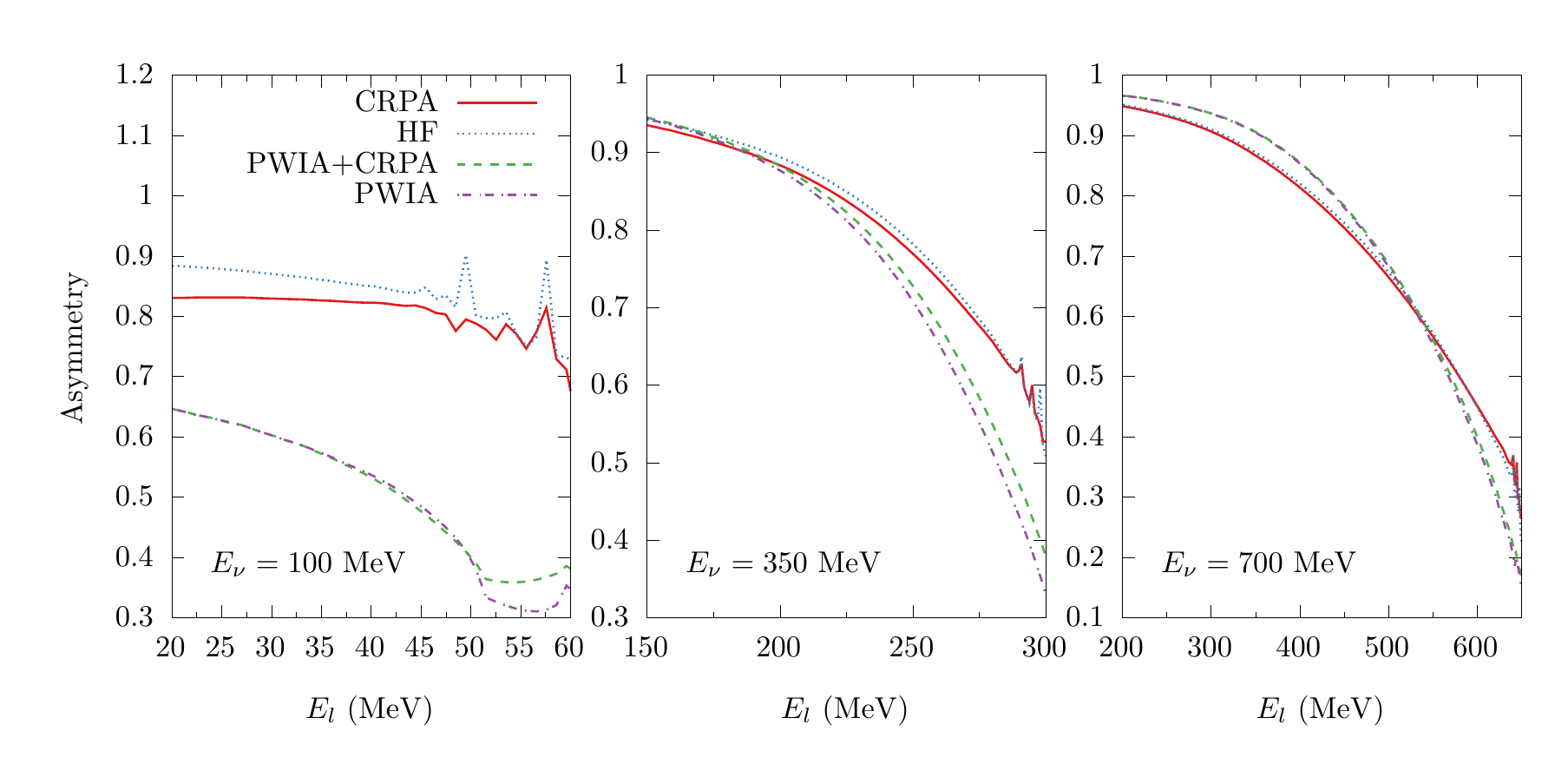}
    \caption{
    Same as in Fig.~\ref{fig:simpelAdep} but for ${}^{40}$Ca and for several different models.
} 
    \label{fig:simpelAdepPW}
\end{figure}


Figure~\ref{fig:simpel} compares charged-current cross section results for electron neutrinos and antineutrinos off a $^{12}$C target, along with the corresponding neutrino-antineutrino asymmetry for different incoming neutrino energies. The high $E_{e}$ tails of the spectrum correspond to reactions with small energy transfers that are most sensitive to nuclear medium effects.  This region is characterized by a number of peaks and resonances stemming from excitations from individual shells and from the giant resonance. From the lower energy panels at the top it is clear that in this energy regime, RPA correlations are the cause of extra strength, whereas for higher kinematics they provide a suppression of the mean-field cross sections.   For lower incoming energies the asymmetry is considerably smaller (i.e. the difference between neutrino and antineutrino-induced cross sections becomes small) than in the genuine QE regime shown in the bottom panels.  Except for the lowest incoming energies the asymmetry is rather insensitive to differences between RPA and HF strength.  The relativistic Fermi gas (RFG) results that are shown clearly illustrate that whereas a Fermi gas calculation can provide a satisfactory overall description of the cross section strength when the energy transfer is sufficiently large, it fails to capture the nuclear medium effects that dominate cross section for low incoming energy and small energy transfers.

In Figure~\ref{fig:DDIFfixEA} the A-dependence of different responses for neutrino- and antineutrino calculations is studied.  The contribution of longitudinal, transverse and transverse interference responses to the cross sections is compared. Clearly, the larger $^{40}$Ca system results in broader cross section distributions with lower peaks and more pronounced tails~\cite{VanDessel:2017ery}.  As both nuclei are isospin symmetric, the difference between neutrino and antineutrino responses should be due solely to the Coulomb potential which affects the protons, mainly through a shift of the binding energy in both initial and final-state. The effect of the distortion of the lepton wave function was not included in this comparison. 
The Coulomb energy for a sphere with charge Z and radius $1.2A^{1/3}$ is $\approx$~4.5 MeV  and $\approx$~10 MeV for carbon and calcium respectively.  These values are indeed approximately the observed shift in the cross sections, which is around twice as large for calcium compared to oxygen. This effect decreases for larger momentum transfers.

In Figs.~\ref{fig:simpelAdep} and \ref{fig:simpelAdepPW} we study the A-dependence and the model-dependence of the asymmetry in  more detail.
We restrict the range of lepton energies to the region in which both the neutrino and antineutrino cross sections have an appreciable size.
Figure~\ref{fig:simpelAdep} shows the asymmetry obtained in CRPA for different nuclei. Apart from the even-even nuclei, ${}^{48}$Ca is also included, for which we take the asymmetry per active nucleon, i.e. dividing the neutrino and anti-neutrino cross sections by $28$ and $20$ respectively.
Apart from apparent deviations for small incoming energies, the asymmetry for the symmetric nuclei is practically independent of the nucleus.
This can be understood from a scaling argument. In a calculation with a single general scaling function (the RFG in particular), one finds that the asymmetry is practically independent of the Fermi-momentum, granted that the Fermi momentum is the same in neutrino and anti-neutrino cross sections. However the asymmetry does still depend on the general shape of the scaling function, and to a smaller extent on the energy shift that is used.
The agreement between the asymmetries for the different even-even nuclei is hence expected, as the differences between the proton and neutron single-particle states in these cases are relatively small.
For the asymmetric ${}^{48}$Ca this is no longer the case, hence it is the odd-one-out.
The comparison at $100~\mathrm{MeV}$ shows that these clear trends disappear in the low-energy region, where the influence of nuclear medium effects and binding-energy differences in particular is larger.

Figure~\ref{fig:simpelAdepPW} examines the model-dependence of these results by comparing the asymmetry for ${}^{40}$Ca obtained with different models.
The PWIA results are obtained by neglecting the distortion of the outgoing nucleon wave function, i.e. describing the outgoing partial waves by spherical Bessel functions, while the rest of the procedure is exactly the same as in the distorted wave calculations. 
One sees that the CRPA and HF cross sections give the same result for the asymmetry, but these distorted wave calculations differ appreciably from the plane-wave results even for the higher energies.

\section{Comparison with data}

\begin{figure}
    \includegraphics[width=\textwidth]{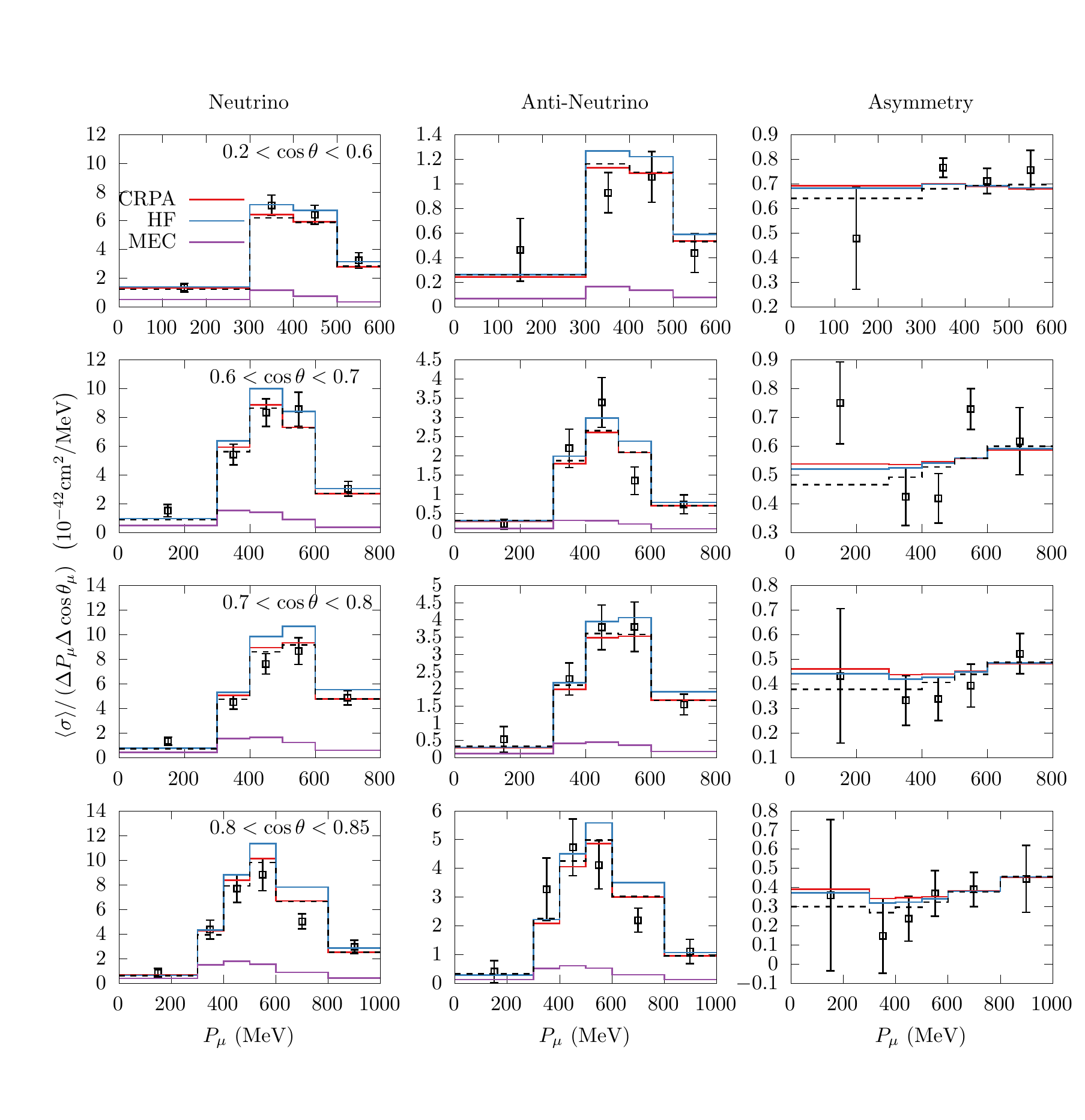}
    \caption{Comparison between HF, CRPA  flux-folded  cross section  and T2K data for neutrinos (left column), antineutrinos (middle), and the asymmetry (right), where each row shows a different range for the lepton scattering angle. 
    The data is from~\cite{T2K:nuanu}.
    The MEC contribution is from Ref~\cite{SuSAMEC}, and is added to the CRPA and HF results.
    The dashed lines correspond to the CRPA results that neglect Coulomb effects and for which the average of the neutrino and antineutrino responses are used (see text for details).
}
    \label{fig:T2K}
\end{figure}
\begin{figure}
    \includegraphics[width=\textwidth]{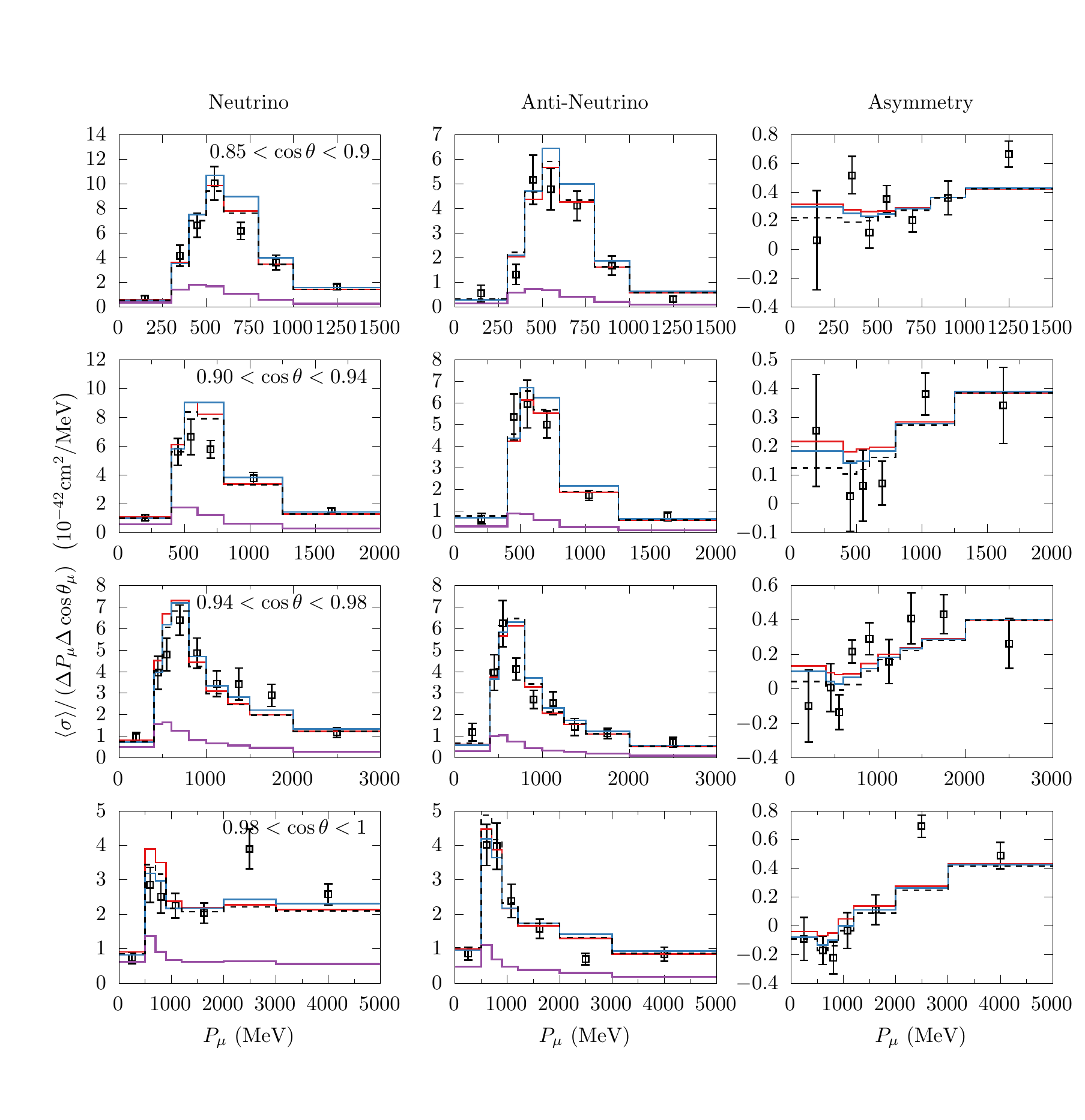}
    \caption{Comparison between HF, CRPA  flux-folded  cross section  and T2K data for neutrinos (left column), antineutrinos (middle), and the asymmetry (right) for forward lepton scattering angles. 
   The data was taken from~\cite{T2K:nuanu}.
    The MEC contribution is from Ref~\cite{SuSAMEC}, and is added to the CRPA and HF results.
    The dashed lines correspond to the CRPA results that neglect Coulomb effects and for which the average of the neutrino and antineutrino responses are used (see text for details).
}
    \label{fig:T2Kbins2}
\end{figure}
T2K has performed a measurement of neutrino and antineutrino cross sections and published  results for the asymmetry,  in terms of the double differential cross section as function of lepton kinematics \cite{T2K:nuanu}.
The comparison to this data is shown in Figs.~\ref{fig:T2K}~and~\ref{fig:T2Kbins2}.
The target is reported to be composed of  87 percent carbon, 7.4 percent hydrogen, 3.7 percent oxygen and 2.7 percent heavier elements (silicon and titanium) by mass.
These mass fractions imply that there is approximately 1 hydrogen particle per carbon nucleus, for  this target composition we calculate cross sections. 
Adding the  oxygen and heavier nuclei contributions has only minor effects on the resulting cross sections per nucleon.
For the antineutrino-hydrogen cross section we only take the charge-changing elastic interaction into account.

To make the comparison complete, we add the SuSav2-MEC calculation of Ref.~\cite{SuSAMEC} for carbon obtained from the inclusive response table included in NuWro~\cite{NuWro:FSI,NuWroSITE}.
The MEC calculation was performed in a symmetric relativistic Fermi gas model, hence the neutrino and antineutrino cross sections are obtained by simply changing the sign of the vector-axial interference term.
We do however also add the effect of the Coulomb potential of the nucleus on the outgoing lepton to the MEC contribution as described above. 
We show the HF and CRPA calculations where no cut-off is used in the residual interaction in the CRPA, as seen in the comparison to electron scattering data these can be interpreted as a  conservative upper (HF) and lower bound (CRPA) for the quasielastic RPA cross section.
Both the HF and CRPA calculations include the appropriate differences in binding energy and the Coulomb potential for the final-state nucleon as discussed above in addition to the effect of the nuclear charge on the final state lepton using the MEMA approach.

With the dashed line, we also show the isospin averaged result obtained in the CRPA calculation, i.e. the result where the average is taken of the charge-changing response on protons and neutrons in carbon and the Coulomb distortion of the outgoing muon is neglected.
This isospin averaged result is very close to the  baseline CRPA result, the main differences that can be seen in some energy bins stem from neglecting the Coulomb energy of the final state lepton rather than the differences in the nuclear response which average out, hence showing that for the carbon nucleus the integrated neutrino-antineutrino cross sections are practically completely related by isospin symmetry as would be expected.

We see that while the HF and CRPA approaches differ in magnitude, with the HF results always being the larger ones except in the most forward bins, the asymmetry is again found to be very similar for both models. This confirms the general assumption that in the asymmetry most uncertainties and medium effects will cancel out, allowing one to use  the asymmetry  to gauge the contribution of other interaction mechanisms as e.g.  MECs as stated in the T2K analysis~\cite{T2K:nuanu}.
We do however note that the asymmetry in the low momentum bins is affected more significantly by Coulomb effects than the individual cross sections, and is moreover quite strongly affected by the hydrogen contribution, for which additional mechanisms beyond the charge-changing elastic cross section included here might contribute.

\section{Comparison with ab-initio calculations}

\begin{figure}
    \includegraphics[width=\textwidth]{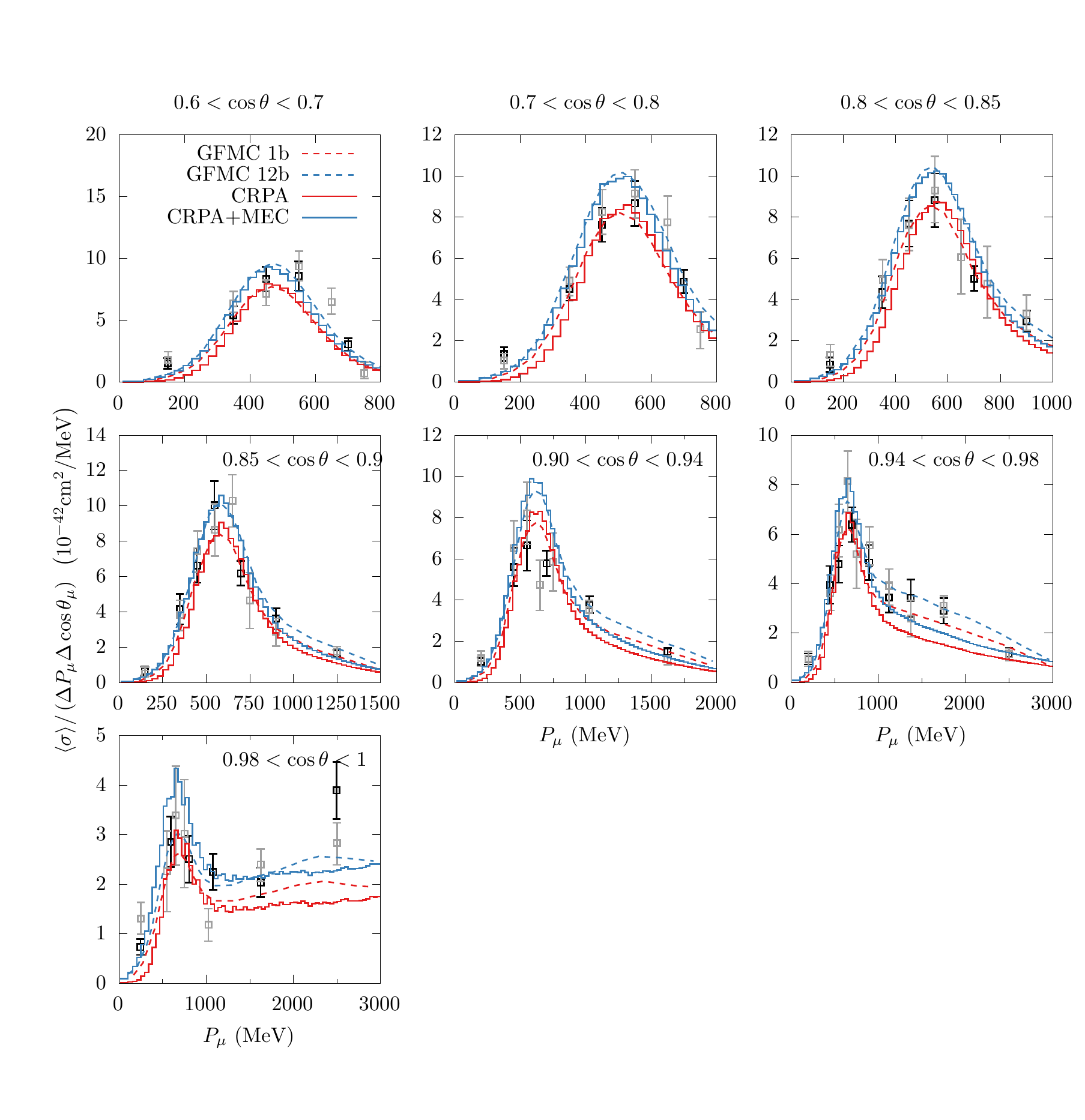}
    \caption{Comparison of CRPA predictions with GFMC results and  T2K neutrino data.  The black data points correspond to the recent T2K dataset \cite{T2K:nuanu} shown in the previous figure. The dark-grey data points correspond to the 2016 dataset, as used in the original GFMC paper \cite{T2K:Inclnumu}.
    Solid blue lines represent the CRPA+SuSAv2-MEC results.
    }
    \label{fig:T2KGFMCCRPA}
\end{figure}

Recently, ab-initio results for inclusive flux-folded neutrino-nucleus cross sections became available \cite{Lovato:PRX}. 
Calculated using quantum Monte-Carlo techniques, ab-initio calculations are in principle exact, up to the accuracy of the interaction that is used and relativistic corrections. 
These results are especially interesting because 1 and 2 body contributions, and the interference between the two, are treated consistently.
Whereas  numerically too time-consuming and lacking the flexibility to provide cross section results for heavier nuclear systems in experimental analyses demanding huge amounts of kinematical settings,  ab-initio results provide an excellent benchmark for testing more efficient calculation schemes.  In this section we provide a comparison between CRPA results and GFMC cross sections, separated in 1- and 1+2-body contributions.  For this comparison, The GFMC results were 
extracted from the originally published plots. All extracted points correspond to the central values given in Ref.~\cite{Lovato:PRX}, the error bands are not shown. In general the error bands become larger for backward scattering angles, a behavior that is particularly outspoken 
for antineutrinos.

Confronting our CRPA calculations with the ab-initio results and T2K data in Fig.~\ref{fig:T2KGFMCCRPA} shows a remarkably good agreement between CRPA results and the 1-body contribution of the GFMC results, adding SuSav2 MEC contributions to the RPA cross sections shows an excellent agreement with the GFMC 1b+2b results.
Both models also agree very well with the data.  Only for the highest $P_{\mu}$ bins, a moderate divergence between both models appears.

\begin{figure}
    \includegraphics[width=\textwidth]{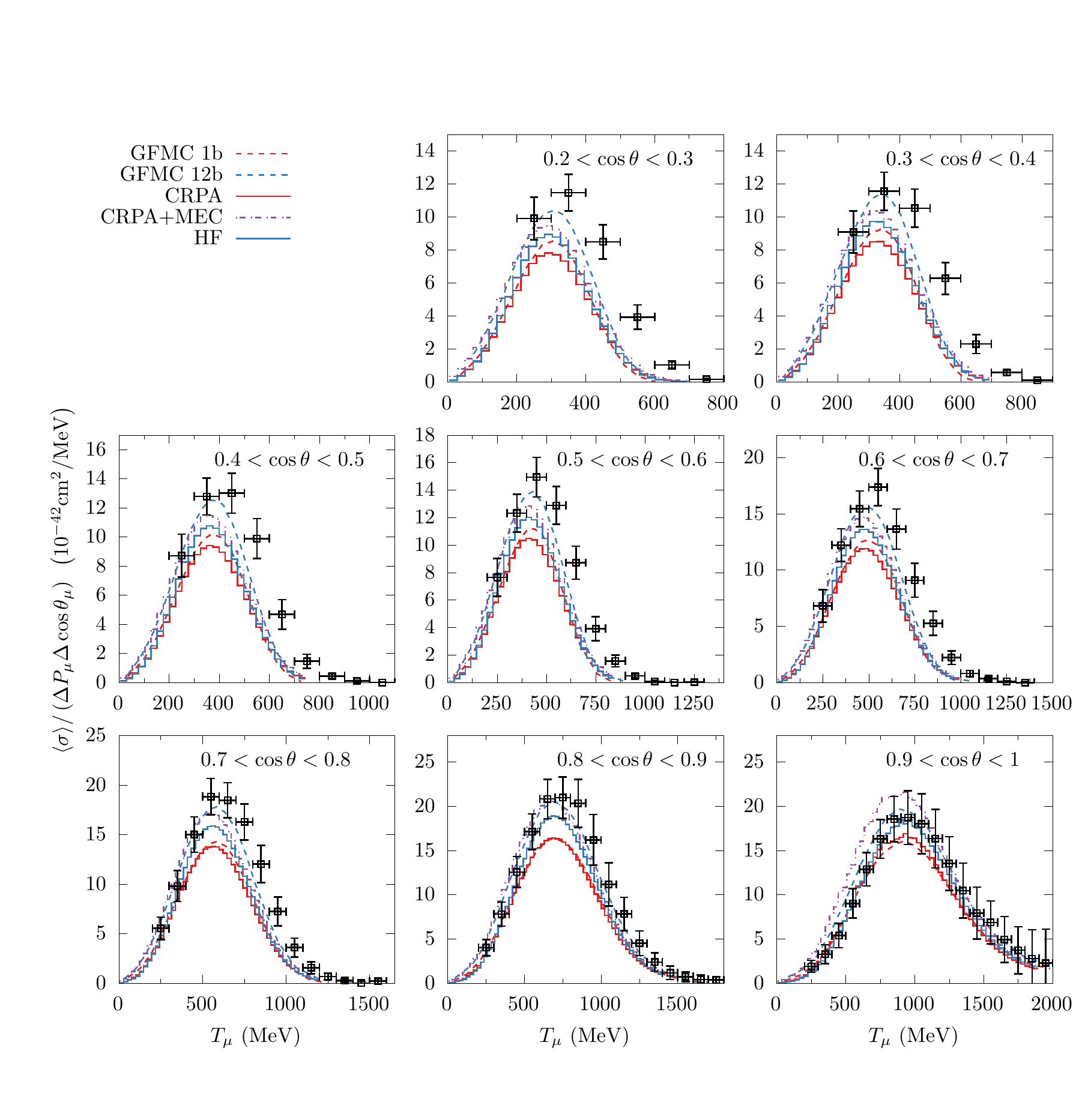}
    \caption{MiniBooNE flux folded results for neutrino scattering off carbon and comparison with CRPA and GFMC calculations. The data is from Ref.~\cite{MB:QECS}}. The dash-dotted purple lines show the result when the SuSAv2-MEC model of Ref.~\cite{SuSAMEC} is added to the CRPA cross section.
    \label{fig:MBGFMCnu}
\end{figure}

\begin{figure}
    \includegraphics[width=\textwidth]{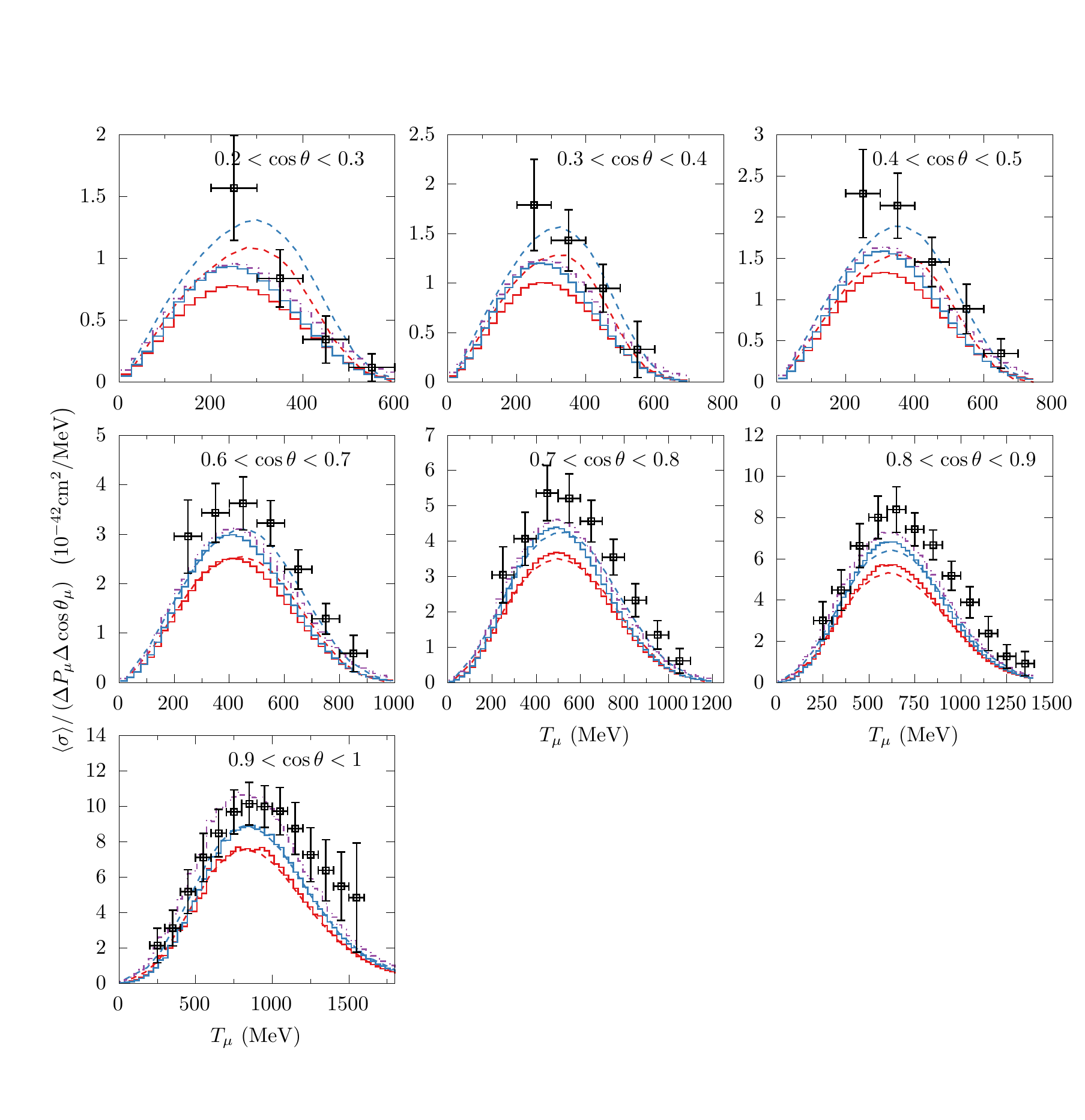}
    \caption{The same as in Fig.~\ref{fig:MBGFMCnu} but for antineutrinos, the data corresponds to the ${}^{12}$C dataset from Ref.~\cite{MB:CCQEanti}. The dash-dotted purple lines show the result when the SuSAv2-MEC model of Ref.~\cite{SuSAMEC} is added to the CRPA cross section.}
    \label{fig:MBGFMCanu}
\end{figure}

For MiniBooNE, the comparison with GFMC results  in Figs.~\ref{fig:MBGFMCnu} and \ref{fig:MBGFMCanu} shows that the CRPA result 
matches the GFMC one-body result exactly in the most forward bins for neutrinos and antineutrinos.  This quasi-perfect agreement slightly deteriorates when considering more backward scattering angles, especially for antineutrinos.
Apart from the fact that the underlying nuclear models are constructed on vastly different foundations such that agreement is not a priori expected, one can pinpoint several possible underlying reasons for this disagreement.  
A first cause finds its origin in the Skyrme parametrization. 
 As discussed above and found in the comparison to inclusive electron scattering data, the Skyrme residual interaction in the CRPA calculation is of zero-range, in practice meaning that the reduction of the cross section found in CRPA becomes too strong at large values of $Q^2$. As in the comparison to T2K data in the previous paragraph we did not  use the phenomenological cut-off in the residual force here. When the residual interaction is turned off completely, one obtains the Hartree-Fock results, while when it is fully included one obtains the  CRPA results shown in the figures.
    As established in the comparison to electron scattering data the full residual force should be included in the forward scattering region (at low $q$), while for more backward angles (high-$q$) where transverse contributions become more dominant, a cut-off is found to be better supported by the $(e,e')$ data, a conclusion that is corroborated by the neutrino results shown here.
    It is clear that whereas the bare mean-field HF prediction tends to overestimate the GFMC results, switching on long-range correlations with the CRPA correction almost perfectly agrees with the ab-initio 1b results, especially for forward bins.  This demonstrates that for these kinematics, the set of correlations included in the CRPA i.e. the expansion of RPA ph-bubbles  to all orders, suffices to  match the 1-body result of the slightly different but more extensive set of diagrams included in the ab-initio calculation.

    Comparing the flux-folded CRPA to the HF results, one sees that the difference between both is rather modest for neutrino-induced interactions, but is more significant for antineutrino-induced interactions.
    This effects finds its origin in the fact that RPA effects reduce the longitudinal response more strongly than the transverse ones as shown in Figs.~\ref{fig:RL} and \ref{fig:RT}, and that, due to the transverse vector-axial interference term entering with a negative sign in antineutrino interactions, the antineutrino cross section receives a relatively stronger longitudinal contribution.
    Taking this into account, one sees that the one-body contribution in the GFMC results is consistent with the HF/CRPA bounds for the neutrino-induced interactions.
  It is moreover interesting to point out that  the GFMC results shown here correspond to the central values reported in Ref.~\cite{Lovato:PRX}.
    In the GFMC calculations however uncertainties due to the underlying nucleon-nucleon interaction are taken into account as shown by error bands~\cite{Lovato:PRX}.
    The size of these uncertainties generally increases for backward scattering angles, and while rather modest in the neutrino results, they becomes quite large in the antineutrino case for backward scattering bins and for large $T_\mu$.
  Finally, we would like to mention that relativistic effects  become more important at higher $q$, and that a non-relativistic reduction of the current in both the GFMC and HF-CRPA approaches, can hence be expected to run into its limits for backward scattering first.

Remarkably, the agreement between the two model predictions does not go hand in hand with an equally accurate description of the data.  For neutrinos, the high $T_{\mu}$ tail of the data tends to be substantially larger than the predictions, especially for more backward scattering and even when 2-body strength is included.  The discrepancy contrasts with the very satisfying agreement that was found for T2K data.  For antineutrinos on the other hand, the underestimation of the data is present over a broader kinematic range but disappears for backward scattering and high $T_{\mu}$.

\begin{figure}
    \includegraphics[width=\textwidth]{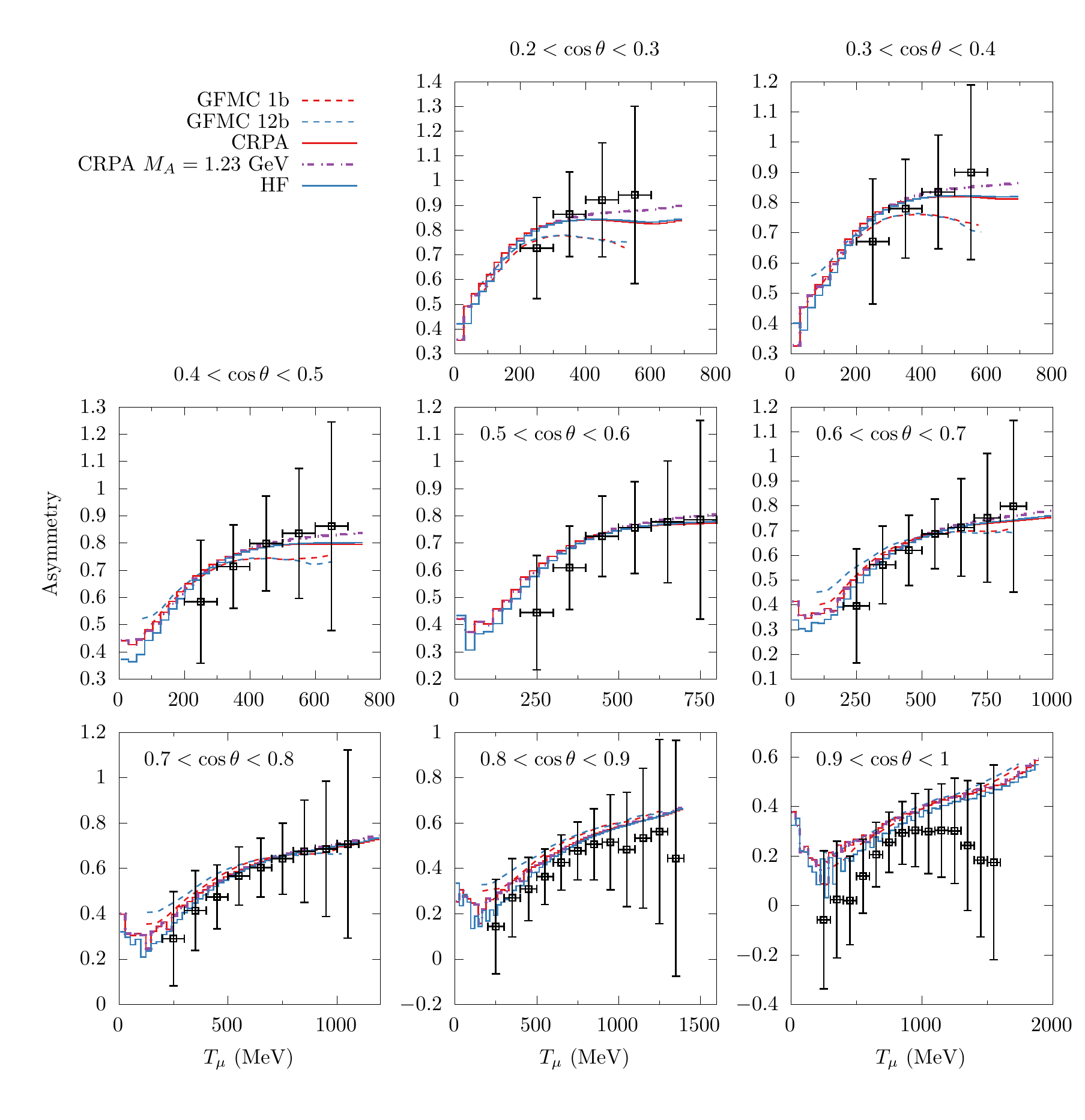}
    \caption{Results for the asymmetry for MiniBooNE.  The asymmetry is computed from the reported neutrino and anti-neutrino data of Refs.~\cite{MB:QECS,MB:CCQEanti}. Error bars for the asymmetry are obtained by combining the individual neutrino and antineutrino uncertainties in quadrature when taking the sum and difference.}
    \label{fig:MBGFMCasym}
\end{figure}
The comparison between the CRPA and GFMC  predictions for the neutrino-antineutrino asymmetry is shown in Fig.~\ref{fig:MBGFMCasym}.
We find very similar results for the GFMC and HF-CRPA results for the most forward scattering bins, as could be expected as they yield practically the same cross section strength in this region.
Once again one finds that the HF and CRPA results give essentially the same result for the asymmetry over the full kinematic range covered by the data, although here the magnitudes of neutrino and antineutrino interactions are different.
Additionally one can observe  that also the one-body and one-plus-two body GFMC cross sections give nearly the same result for the asymmetry, illustrating again that the asymmetry is indeed a robust quantity with little sensitivity to medium effects.
The differences between the HF-CRPA  and the GFMC results appear mainly for large $T_\mu$ bins and at backward scattering angles, and  hence find their origin in the difference in strength for antineutrino cross sections obtained in both models. For these kinematics the uncertainties on the GFMC results, not shown here, are however larger. 

While the comparison of HF/CRPA cross sections with the neutrino and antineutrino data in MiniBooNE tends to result in an underestimation of the high-$T_\mu$ tail, the results for the asymmetry do follow the trend of the datapoints very well in this region.
In Fig.~\ref{fig:MBGFMCasym} we also included the CRPA results where the axial mass is increased to a value of $1.23~\mathrm{GeV}$. The asymmetry is rather insensitive to such a change except in the most backward bins.
Because we extract the asymmetry from the published neutrino and antineutrino cross sections, adding the reported errors in quadrature, the shown errorbars are likely to be an overestimation. 
It can be expected that e.g. flux normalization uncertainties cancel to a large extent when ratios between the cross section are taken.
Under the assumption that the error would be much reduced in a fully consistent analysis, it is interesting to note that the most obvious discrepancy between the models and the asymmetry data is actually found in the most forward scattering bin.

\section{Summary}
We have performed an extensive comparison of neutrino and antineutrino-induced cross sections relevant for current and future oscillation experiments.
We paid special attention to effects inflicted by the nuclear medium on responses. Confronting ab-inito predictions with our results provided by more effective calculation schemes, we find an excellent agreement when considering corresponding interaction mechanisms and nuclear medium ingredients.
We find a good agreement with T2K neutrino data, but for MiniBooNE some remarkable tensions between data and predictions appear.
The neutrino-antineutrino cross section asymmetry was found to be a robust quantity with relatively little influence of medium corrections except at very small energy transfers.

\begin{acknowledgement}
The authors thank Stephen Dolan for information regarding the T2K data.
This work was supported by the Research Foundation Flanders (FWO-Flanders),
A.N. acknowledges the support of the Swiss confederation through ESKAS nr 2020.0004.
\end{acknowledgement}

\section*{Author's contributions}
A.N. performed the calculations and made the figures, N.J. wrote the manuscript, both authors revised the manuscript.

\bibliographystyle{apsrev4-1.bst}
\bibliography{BiblioG}

\end{document}